\documentclass{jfm}
\usepackage{graphicx}
\usepackage{epstopdf, epsfig}
\usepackage{bm}
\usepackage{amsmath,amsfonts}
\usepackage{hyperref} 
\usepackage{caption}
\usepackage{subcaption}
\usepackage{csquotes}
\usepackage{xcolor}
\usepackage{MnSymbol}

\title{Effect of geometric disorder on chaotic viscoelastic porous media flows}

\author{ A. Chauhan, S. Gupta, C. Sasmal\aff{1}
  \corresp{\email{csasmal@iitrpr.ac.in}},
  }

\affiliation{\aff{1}Soft Matter Engineering and Microfluidics Lab, Department of Chemical Engineering, Indian Institute of Technology
Ropar, Rupnagar, India-140001}

\begin{document}
\maketitle
\begin{abstract}
The flow of viscoelastic fluids in porous media is encountered in many practical applications, such as in the enhanced oil recovery process or in the groundwater remediation. Once the flow rate exceeds a critical value in such flows, an elastic instability with fluctuating flow field is observed, which ultimately transits to a more chaotic and turbulence-like flow structure as the flow rate further increases. In a recent study, it has been experimentally shown that this chaotic flow behaviour of viscoelastic fluids can be suppressed by increasing the geometric disorder in a model porous media consisting of a microchannel with several micropillars placed in it. However, the present numerical study demonstrates that this is not always true. We show that it depends on the initial arrangement of the micropillars for mimicking the porous media. In particular, we find that for an initial ordered and aligned configuration of the micropillars, the introduction of geometric order actually increases the chaotic flow dynamics as opposed to that seen for an initial ordered and staggered configuration of the micropillars. We suggest that this chaotic flow behaviour actually depends on the number of the stagnation points revealed to the flow field where maximum stretching of the viscoelastic microstructure happens. Our findings and explanation are perfectly in line with that observed and provided in a more recent experimental study.
\end{abstract}


\section{\label{Intro}Introduction}
Viscoelastic fluids are a class of fluids that exhibit both viscous and elastic responses under applied deformation. These fluids, such as polymer solutions or melts, foams, emulsions, etc., are widely used in scores of industrial settings ranging from chemical to healthcare and cosmetic industries~\citep{chhabra2011non}. The elastic response of these fluids, arising from the relaxation mechanism of the constituent microstructure, is often dictated by the dimensionless Weissenberg number, defined as $Wi = \lambda \dot{\gamma}$, where $\lambda$ is the fluid relaxation time and $\dot{\gamma}$ is the strain rate~\citep{bird1987dynamics1}. In the flows of these viscoelastic fluids,  when this Weissenberg number exceeds a critical value, a chaotic and turbulent-like flow behaviour is often observed. Due to the fluid elasticity driven nature of this turbulence, it is termed as the 'elastic turbulence', which happens at negligible inertia and/or $Re << 1$~\citep{groisman2000elastic,steinberg2021elastic}. This is in contrast to regular turbulence which is driven by fluid inertia. Before the transition of the elastic turbulence, an elastic instability emerges in the system, which is the precursor of this turbulence. This instability is caused due to the generation of high elastic tensile stresses along a curved streamline, resulting from the stretching of the polymer molecules~\citep{pakdel1996elastic,mckinley1996rheological}. 

Therefore, the onset of both the elastic instability and elastic turbulence is highly dependent on the geometry in which the flow happens. One such example is the flow through porous media wherein highly tortuous flow paths and/or curved streamlines of the fluids can be seen. Furthermore, in porous media, different extents of shear and extensional flow-dominated regions are present, which cause different extents of stretching of the polymer molecules in the flow field. Therefore, the resulting macroscopic flow behaviour becomes very complex and rich in physics. It is imperative to understand this complex flow behavior because of its theoretical importance and its better and efficient use in many practical applications, such as enhanced oil recovery, soil remediation, filtration, etc. As a result, an extensive amount of studies have been carried out to understand the flow behaviour of viscoelastic fluids in a porous media~\citep{sochi2010non}.       

To understand the flow behaviour at the microscale, a model porous media consisting of a microchannel with many micropillars placed in it, is very often used~\citep{browne2020pore}. Many complex and interesting viscoelastic flow behaviour were revealed both experimentally and numerically using such model porous geometry. For instance, De et al.~\citep{de2017lane} found the formation of preferential paths or lanes during the flow of these viscoelastic fluids through such a model porous media wherein the micropillars were arranged in a staggered manner. A spatiotemporal  variation of these paths was also observed as the Weissenberg number gradually increased. Recently, Walkama et al.~\citep{walkama2020disorder} noticed that the flow became chaotic and turbulent-like in this staggered geometry once the Weissenberg number exceeded a critical value. However, they observed that this chaos in the flow is suppressed as some random geometric disorder is introduced in the initial staggered (and ordered) arrangements of the micropillars. The reason behind this, as explained by Walkama et al., is that the introduction of geometric disorder in the ordered porous geometry opens up more preferential paths or lanes, thereby increasing shear flow regions and reducing extensional flow dominated regions. This results in less stretching of the polymer molecules, which in turn inhibits the chaotic fluctuations. However, more recently, Haward et al.~\citep{haward2021stagnation} demonstrated that the geometric disorder does not always suppress these chaotic fluctuations. It depends on the initial configuration of the micropillars placed in the microchannel to mimic the porous media. To prove this, they did experiments with a model porous media wherein the micropillars are initially placed both in aligned and staggered configurations. For the case of aligned configuration, they observed that the introduction of geometric disorder into this configuration increases the chaotic fluctuations as opposed to that seen in a staggered configuration. Therefore, they suggested that it is not always the geometric disorder but the number of stagnation points present in the system (which are formed at each pillar's front and back surfaces and induce significant stretching of polymer molecules), which ultimately controls these chaotic fluctuations in porous media.              

Therefore, it can be seen that the experimental studies performed by Walkama et al. and Haward et al. presented a contradict conclusion. However, both the studies agreed that the chaotic fluctuations in the viscoelastic porous media flows result from the stretching of the viscoelastic microstructure in the flow field. This study investigates the same problem through numerical simulations and provides an in-depth understanding of the underlying physics associated with the viscoelastic porous media flows. In particular, we study the effect of the geometric disorder on the chaotic flow dynamics of viscoelastic fluids in a porous media consisting of micropillars.         

\begin{figure}
    \centering
    \includegraphics[trim=0cm 21cm 0cm 2cm,clip,width=13cm]{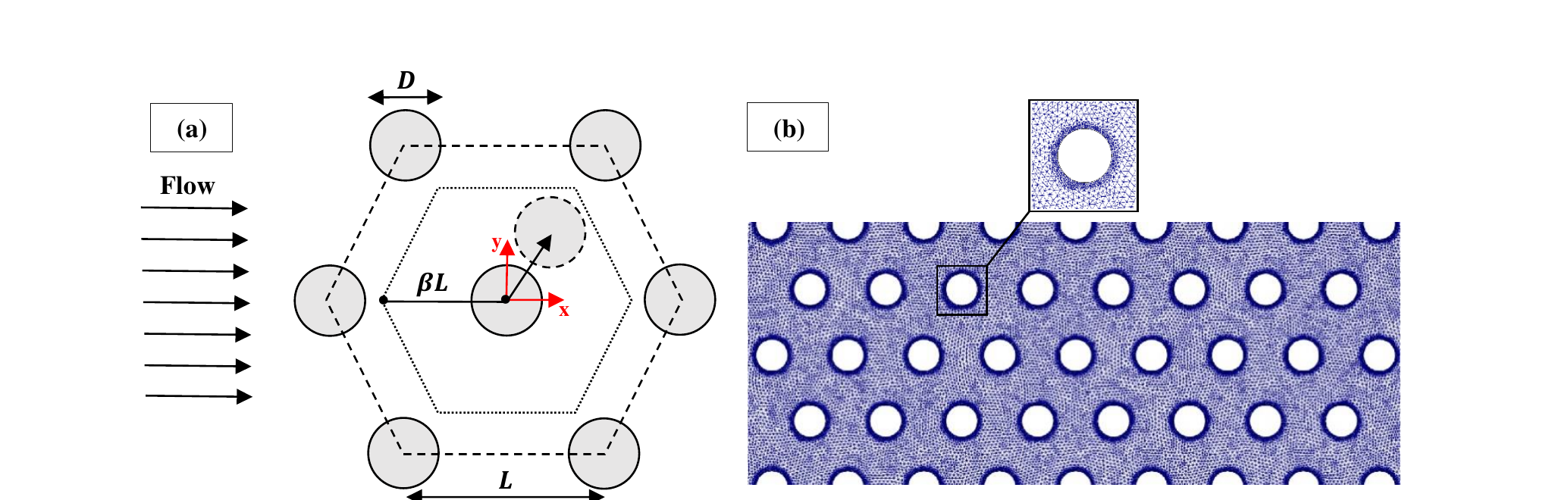}
    \caption{(a) Schematic of the present problem. Here a unit representative cell of the aligned configuration of the micropillars is shown (b) A typical mesh structure used in the present study with a zoom view around a micropillar.}
    \label{fig:fig1}
\end{figure}

\section{\label{ProbFor}Problem description, governing equations and numerical details}
In this study, the flow of viscoelastic fluids through a model porous media, consisting of a microchannel with many circular micropillars present in it, is studied in the creeping flow regime. The micropillars (of diameter $D$) are initially placed in the microchannel as hexagonal arrays with an aligned configuration. A unit representative cell for the arrangement is shown in figure~\ref{fig:fig1}(a). The geometric disorder in such model porous media is introduced by the random displacement of each micropillar within a hexagon of circumradius $\beta L$~\citep{walkama2020disorder}, where $\beta$ is the disorder parameter and $L$ is the separation distance between any two micropillars in their initial aligned configuration, as schematically shown in figure~\ref{fig:fig1}(a). In this study, a polymer solution is used as a model viscoelastic fluid, which can be made by adding a minute amount of polymers into a Newtonian solvent like water. The rheological behaviour of the resulting polymeric fluid is realized based on the FENE-P (finitely extensible non-linear elastic spring with the Peterlin's approximation) constitutive model~\citep{bird1995constitutive,bird1987dynamics}. In this model, each polymer molecule is assumed to be a dumbbell with two beads connected by a finitely extensible elastic spring. It is extensively used for the viscoelastic fluid simulations, particularly for the turbulent viscoelastic flows~\citep{sureshkumar1997direct,richter2012effects}. The polymer solution is also assumed to be incompressible in nature. Under these circumstances, the present flow will be governed by the following equations\newline
\newline
Continuity equation
\begin{equation}
   \boldsymbol{\nabla} \cdot \mathbf{u}
\end{equation}
Cauchy momentum equation 
\begin{equation}
    \rho \left ( \frac{\partial \mathbf{u}}{\partial t} + \mathbf{u} \cdot \boldsymbol{\nabla} \mathbf{u} \right) = - \boldsymbol{\nabla} p + \boldsymbol{\nabla} \cdot \boldsymbol{\tau}
\end{equation}
In the above equations, $\mathbf{u}$ is the velocity vector, $\boldsymbol{\tau}$ is the total extra stress tensor, $\rho$ is the fluid density, $t$ is the time and $p$ is the pressure. The total extra stress tensor is comprised of two components, namely, one from the solvent $(\boldsymbol{\tau}_{s})$ and another from the polymers $(\boldsymbol{\tau}_{p})$, i.e., $\boldsymbol{\tau} = \boldsymbol{\tau}_{s} + \boldsymbol{\tau}_{p}$. The solvent contribution is given as $\boldsymbol{\tau}_{s} = 2 \eta_{s} \mathbf{D}$ with $\mathbf{D} = \frac{1}{2} \left(  \nabla \mathbf{u} + \nabla \mathbf{u}^{T} \right)$ where $\eta_{s}$ is the solvent viscosity. The polymeric contribution is provided by the following equation as per the FENE-P constitutive relation
\begin{equation}
    \boldsymbol{\tau}_{p} + \frac{\lambda}{f} \overset{\kern0.25em\smalltriangledown}{\boldsymbol{\tau}}_{p} = \frac{a \eta_{p}}{f} \left( \boldsymbol{\nabla} \mathbf{u} + \boldsymbol{\nabla} \mathbf{u}^{T} \right) - \frac{\text{D}}{\text{D}t} \left(  \frac{1}{f}\right) \left[ \lambda \boldsymbol{\tau}_{p} + a \eta_{p} \mathbf{I} \right]
\end{equation}
where $f = \frac{L^{2} + \frac{\lambda}{a \eta_{p}}\text{tr}(\boldsymbol{\tau})}{L^{2} - 3}$, $a = \frac{L^{2}}{L^{2}-3}$, $\eta_{p}$ is the zero-shear rate polymer viscosity, $\lambda$ is the polymer relaxation time, $L^{2}$ is the polymer extensibility parameter, and $\overset{\kern0.25em\smalltriangledown}{\boldsymbol{\tau}}_{p} = \frac{\partial \boldsymbol{\tau}_{p}}{\partial t} + \mathbf{u} \cdot \boldsymbol{\nabla}\boldsymbol{\tau}_{p} - \boldsymbol{\tau}_{p} \cdot \boldsymbol{\nabla} \mathbf{u} - \boldsymbol{\nabla} \mathbf{u}^{T} \cdot \boldsymbol{\tau}_{p}$ is the upper convected time derivative of $\boldsymbol{\tau}_{p}$. The following dimensionless numbers are introduced in the present study, namely, Reynolds number $\left( Re = \frac{D U_{in} \rho}{\eta_{0}}\right)$, Weissenberg number $\left( Wi = \frac{\lambda U_{in}}{D}\right)$, and polymer viscosity ratio $\left( \beta_{v} = \frac{\eta_{s}}{\eta_{0}} \right)$ where $\eta_{0} = \eta_{s} + \eta_{p}$ is the zero-shear rate viscosity of the polymer solution. Along with these, the polymer extensibility parameter $L^{2}$ is another dimensionless parameter used in this study. The following values of these dimensionless numbers are chosen for the present study: $\beta_{v} = 0.59$, $L^{2} = 500$, $Re = 0$ and $Wi = 0 - 10$. 

The finite volume method-based OpenFOAM toolbox~\citep{wellerOpenFOAM} along with rheoTool (version 5)~\citep{rheoTool}, have been used to solve the aforementioned governing equations, namely, continuity, momentum, and FENE-P viscoelastic constitutive equations. The advective terms in the momentum and constitutive equations were discretized using the high-resolution CUBISTA (Convergent and Universally Bounded Interpolation Scheme for Treatment of Advection) scheme for its improved iterative convergence properties~\citep{alves2003convergent}. The diffusion terms in both the momentum and constitutive equations were discretized using the second-order accurate Gauss linear orthogonal interpolation scheme. All convective terms were discretized using the Gauss linear interpolation scheme.  While the linear systems of the pressure and velocity fields were solved using the Preconditioned Conjugate Solver (PCG) with DIC (Diagonal-based Incomplete Cholesky) preconditioner, the stress fields were solved using the Preconditioned Bi-conjugate Gradient Solver (PBiCG) solver with DILU (Diagonal-based Incomplete LU) preconditioner~\citep{ajiz1984robust}. The pressure-velocity coupling was accomplished using the SIMPLE method, and the log-conformation tensor approach was used to stabilize the numerical solution~\citep{pimenta2017stabilization}. Furthermore, the relative tolerance level for the pressure, velocity, and stress fields was set as 10$^{-10}$. A proper choice of the mesh density is needed for the present results to be free from numerical artifacts without demanding excessive computational resources. In this study, for each value of the disorder parameter $\beta$, three grids, namely, G1, G2, and G3, were created with an increasing number of cells in the computational domain. A triangular-shaped grid is used to discretize the present computational domain with a total number of cells (G2 grid density) in between 131202 and 132219 was found to be sufficient for the present study. This was confirmed by comparing the time-averaged values of the root mean square velocity fluctuations (as well as stress) at three different grid densities, and it was seen that the changes became almost negligible as we moved from grid G2 to G3. A typical grid structure used in the present study is shown in sub-figure~\ref{fig:fig1}(b). Finally, the following boundary conditions have been employed to complete the problem formulation. For the velocity, the standard no-slip boundary condition on all solid walls, a fully developed velocity profile at the channel inlet and a zero gradient at the outlet are used. For the polymeric stress tensor, all the components are linearly extrapolated to zero on solid walls, a zero constant value at the inlet and a zero gradient at the outlet are used.

\section{\label{Ressult}Results and discussion}
\begin{figure}
    \centering
    \includegraphics[trim=0cm 17cm 0cm 2cm,clip,width=13.5cm]{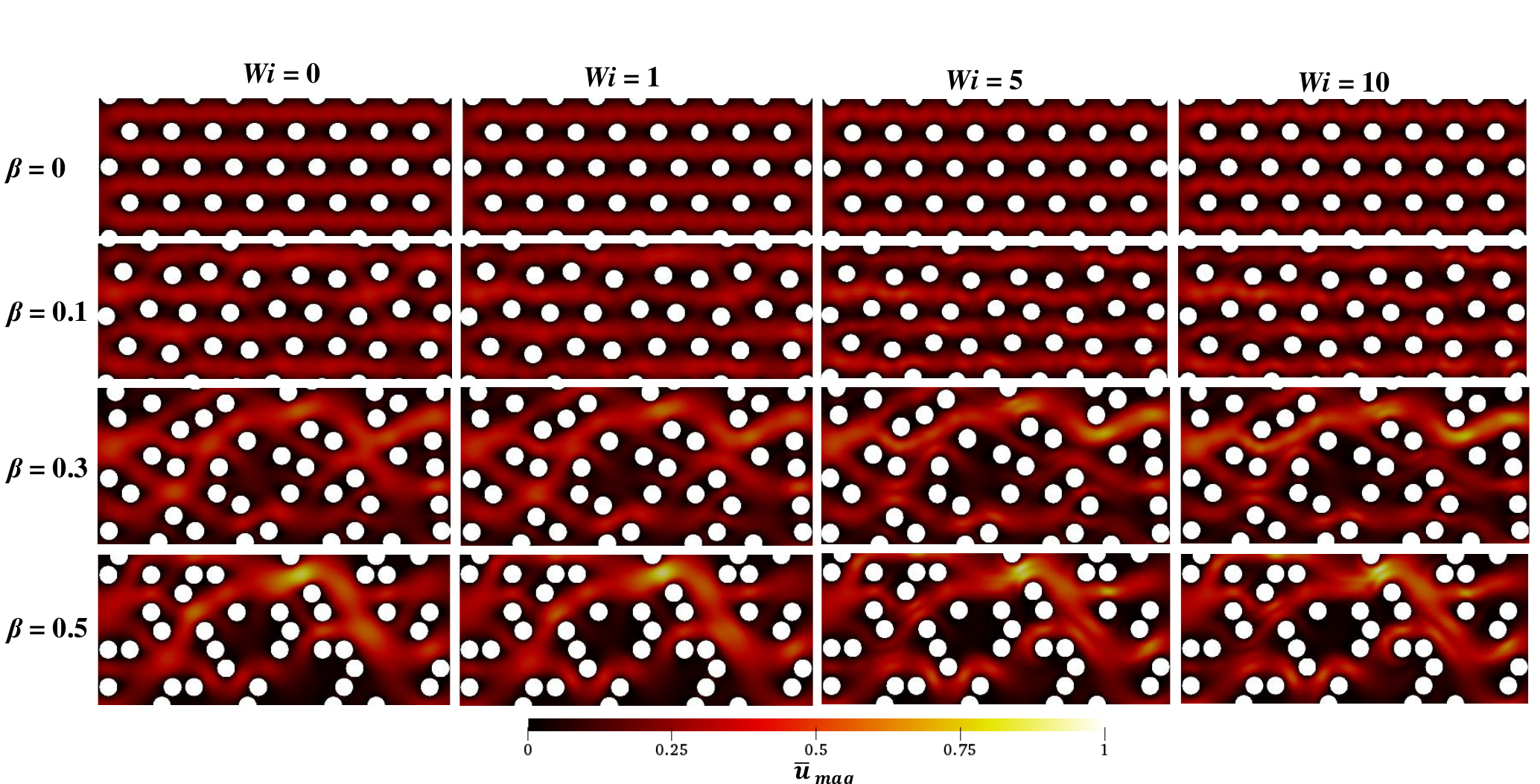}
    \caption{Variation of the time-averaged mean velocity magnitude with the Weissenberg number and disorder parameter.}
    \label{fig:fig2}
\end{figure}

We have carried out simulations over a range of the Weissenberg number in between 0 and 10. We have also included the results for a Newtonian fluid to show the complex flow dynamics associated with the corresponding viscoelastic polymeric fluids under otherwise identical conditions. Simulations have been carried out for aligned configuration of the micropillars with the disorder parameter $\beta$ ranged in between 0 and 0.5. Figure~\ref{fig:fig2} shows the time-averaged velocity magnitude plot $\left( \Bar{u}_{mag} = <\Tilde{u}_{mag}>_{t} \right)$ to show the velocity field inside the porous matrix over a wide range of the Weissenberg number and disorder parameter. For an initial ordered and aligned configuration of the micropillars ($\beta = 0$), one can easily see that the fluid passes through the space in between the two rows of the micropillars, and mostly avoids the space in between the space of two micropillars in any row in the flow direction. Therefore, a dead zone is created in all these regions, which can be clearly evident from a very low value or almost zero value of the velocity magnitude in these regions. This is simply because of the availability of the area in between two rows of the micropillars for the fluid to flow both for Newtonian and viscoelastic fluids. These flow paths are seen to be straight both for Newtonian and viscoelastic fluids with low Weissenberg numbers; however, as the Weissenberg number gradually increases, these become distorted, for instance see the results for the highest value of the Weissenberg number considered in this study, i.e., at $Wi = 10$. 

\begin{figure}
    \centering
    \includegraphics[trim=0.9cm 17cm 1cm 3cm,clip,width=14.3cm]{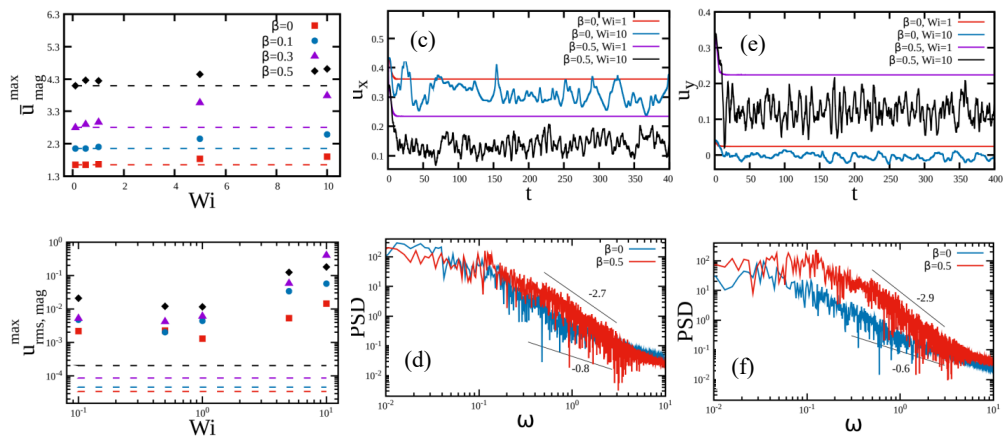}
    \caption{(a) Variation of the maximum value of the time-averaged mean velocity magnitude and (b) the time-averaged root mean square velocity magnitude fluctuations with $Wi$ and $\beta$. Temporal variation of the non-dimensional (c) stream-wise and (e) span-wise velocity components with $Wi$ and $\beta$. The corresponding power spectral density plot of the (d) stream-wise and (f) span-wise velocity component fluctuations at two values of $\beta$, namely, 0 and 0.5.}
    \label{fig:fig3}
\end{figure}

As the geometric disorder is gradually introduced in the ordered aligned configuration of the micropillars, these straight flow paths are naturally destroyed due to the obstruction caused by the micropillars, and some preferential paths or lanes are seen to form through which most of the fluid flows irrespective of the type, i.e., Newtonian or viscoelastic. However, this tendency further increases with the increasing values of both the Weissenberg number and the disorder parameter. This is clearly seen from the time-averaged velocity magnitude plot presented in figure~\ref{fig:fig2}. It is more evident in sub-figure~\ref{fig:fig3}(a) wherein the maximum values of this time-averaged mean velocity magnitude $(\Bar{u}_{mag}^{max})$ are plotted with the Weissenberg number and disorder parameter. It can be seen that the values of $\Bar{u}_{mag}^{max}$ increase both with $Wi$ and $\beta$. The formation of such preferential paths or lanes in viscoelastic fluids has already been seen in earlier experiments~\citep{de2017lane}. 

\begin{figure}
    \centering
    \includegraphics[trim=0cm 16.5cm 0cm 2cm,clip,width=13.5cm]{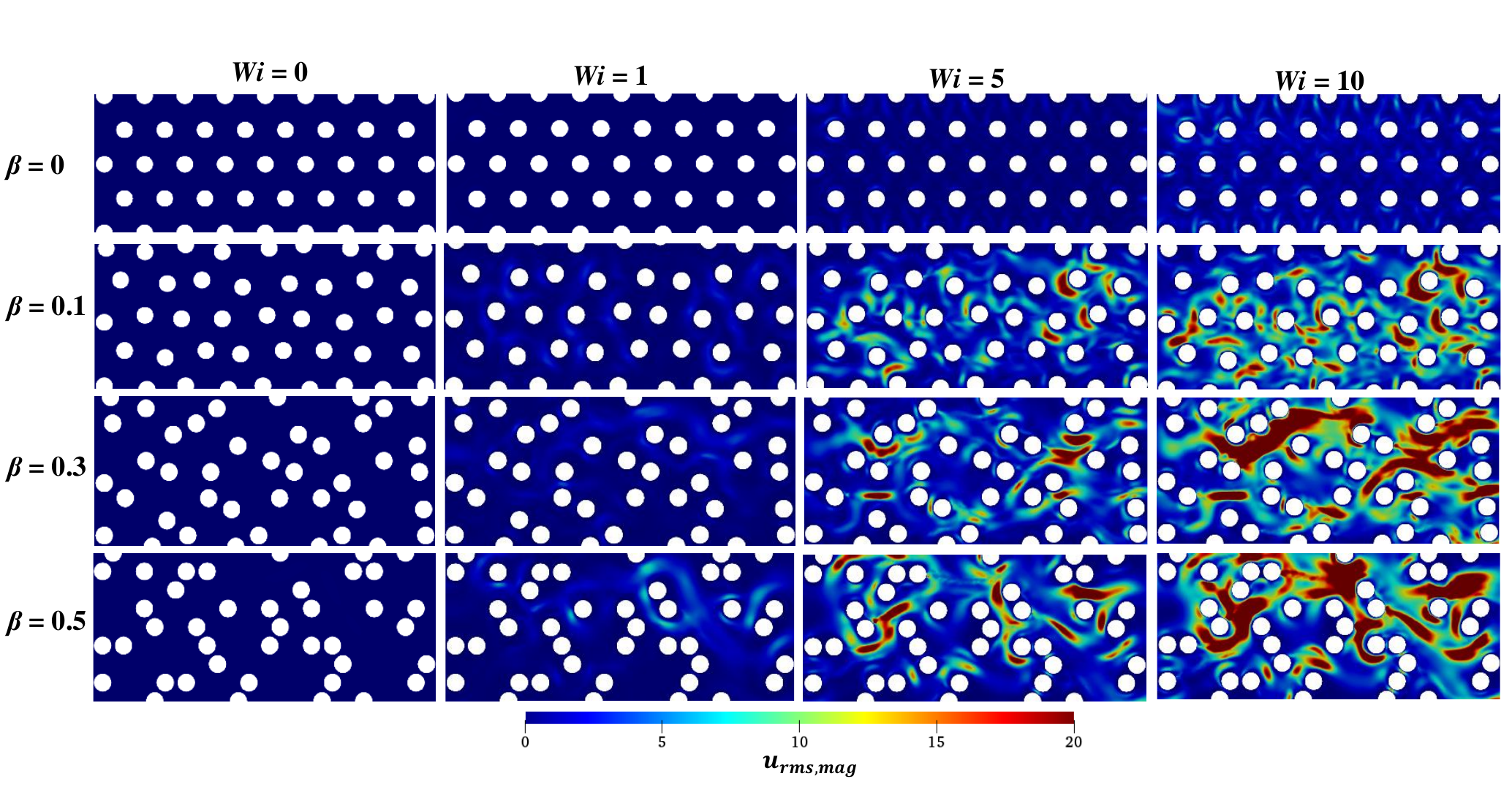}
    \caption{Variation of the root mean square velocity magnitude fluctuations with the Weissenberg number and disorder parameter.}
    \label{fig:fig4}
\end{figure}
The variation of the time-averaged values of the root mean square velocity magnitude fluctuations $\left({u}_{rms,mag} = \sqrt{<(\Tilde{u}_{mag} - \Bar{u}_{mag})^{2}>_{t}} \right)$ with the Weissenberg number and disorder parameter are depicted in figure~\ref{fig:fig4}. It is clearly seen that for a Newtonian fluid $(Wi = 0)$, there is no fluctuation present at all irrespective of the disorder parameter, as it is expected for these fluids in the creeping flow regime. However, as the Weissenberg number gradually increases, ${u}_{rms,mag}$ shows finite values even for the perfectly aligned configuration with $\beta = 0$, see the results at $Wi = 10$. On the other hand, at any fixed value of the Weissenberg number, as the disorder parameter $\beta$ is gradually introduced  in an initial ordered and aligned configuration, one can clearly observe that the velocity fluctuations increase. It becomes more prominent at the highest value of the Weissenberg number ($Wi = 10$) considered in this study. This is more clearly noticeable in sub-figure~\ref{fig:fig3}(b) wherein the maximum values of ${u}_{rms,mag}$ are plotted as a function of the Weissenberg number and disorder parameter. Therefore, it shows that the introduction of geometric disorder actually increases the intensity of the velocity fluctuations in an initial ordered and aligned configuration of the micropillars. This is in contrast to that seen for an initial ordered and staggered configuration of the micropillars by Walkama et al.~\citep{walkama2020disorder} in their experiments. For this latter arrangement, the geometric disorder actually decreased the chaotic and fluctuating behaviour of the flow field. However, our findings are in line with that observed in a more recent experimental study carried out for an initial aligned configuration by Haward et al~\citep{haward2021stagnation}.  

To understand the nature of the chaotic fluctuations in the flow field, we have plotted the temporal variation of both the non-dimensional stream-wise $(u_{x})$ and span-wise $(u_{y})$ velocities at a probe location at two different values of the Weissenberg number (namely, 1 and 10) and disorder parameter (namely, 0 and 0.5) in sub-figure~\ref{fig:fig3}(c) and (e), respectively. At $Wi = 1$, both the velocity components reach a steady value with time irrespective of the disorder parameter. However, as the Weissenberg number increases to 10, both the velocity components show chaotic fluctuations in their temporal variation regardless of the disorder parameter. For perfectly ordered and aligned arrangement $(\beta = 0)$, the fluctuation is seen to be less as compared to that seen for a disordered arrangement $(\beta = 0.5)$, particularly in the temporal variation of the span-wise velocity component (sub-figure~\ref{fig:fig3}(e)). The corresponding power spectral density (PSD) plots of the velocity fluctuations are presented in sub-figures~\ref{fig:fig3}(d) and (f) to characterize the nature of the fluctuations. From both these plots, one can see the excitation of the fluid motion over a wide range of continuum frequencies, which is one of the characteristic features of elastic turbulence. At the same value of the Weissenberg number, the range of the power spectrum is smaller for an ordered aligned configuration than that seen for a disordered one. Therefore, it again suggests that the intensity of fluctuations is more for the latter configuration than that for the former one. A plateau in the power spectrum is seen in the low-frequency range, and at high frequencies, a power-law decay $(\omega^{\alpha})$ is seen in the power spectrum, which covers almost a decade of the frequency range. The fitted values of the power-law exponent $\alpha$ are -2.7 and -2.9 for the stream-wise and span-wise velocity fluctuations, respectively, for a disordered aligned configuration, whereas the corresponding values for an ordered aligned configuration are -0.6 and -0.7, respectively. These values of the power-law exponent clearly suggest the presence of the elastic turbulence for a disordered aligned configuration $(\beta = 0.5)$ but not for an ordered one $(\beta = 0)$~\citep{steinberg2021elastic,haward2021stagnation}.   
    
\begin{figure}
    \centering
    \includegraphics[trim=0cm 16.5cm 1cm 1cm,clip,width=13.3cm]{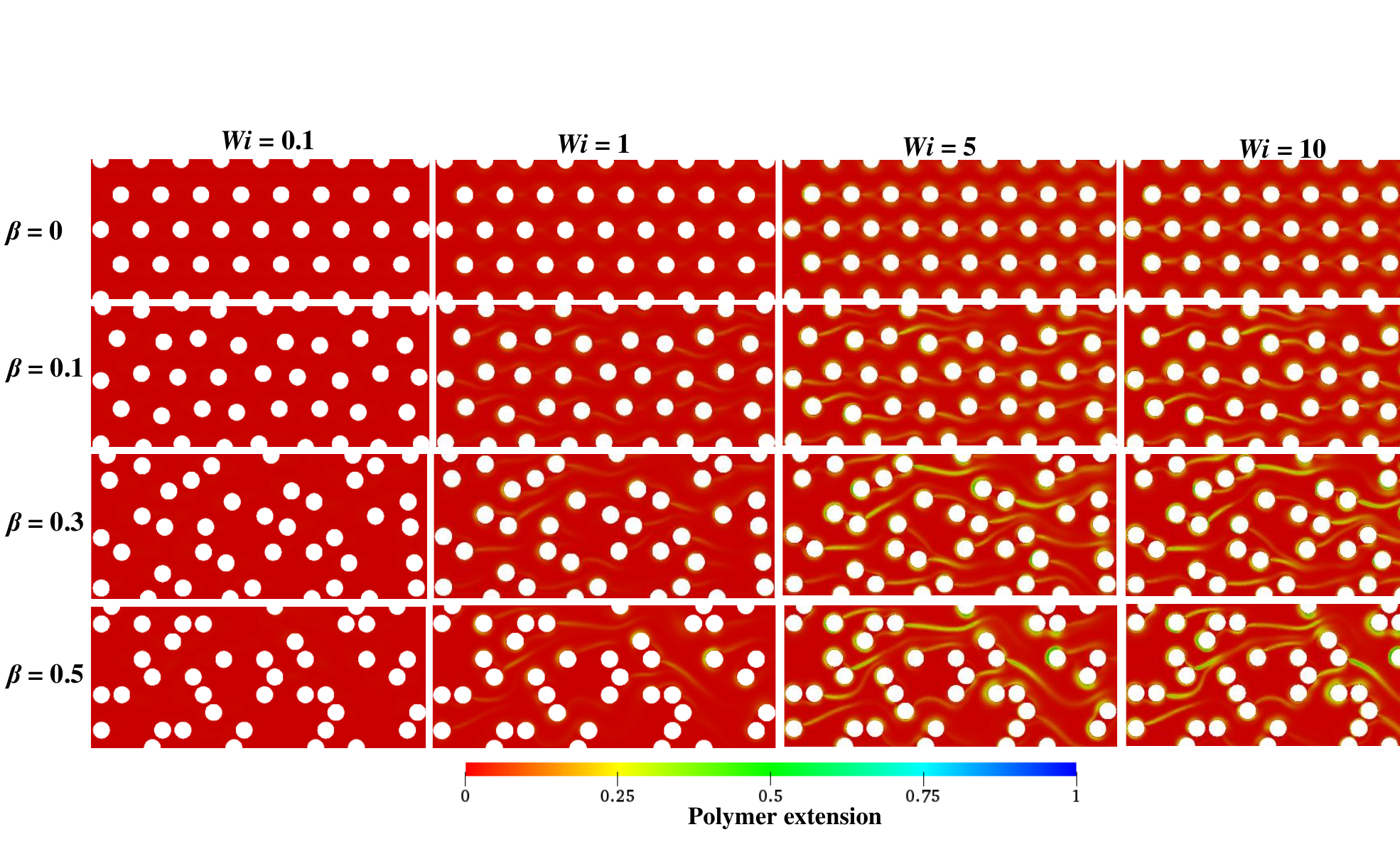}
    \caption{Variation of the polymer extension with the Weissenberg number and disorder parameter.}
    \label{fig:fig5}
\end{figure}

\begin{figure}
    \centering
    \includegraphics[trim=0cm 19cm 1cm 1cm,clip,width=13.4cm]{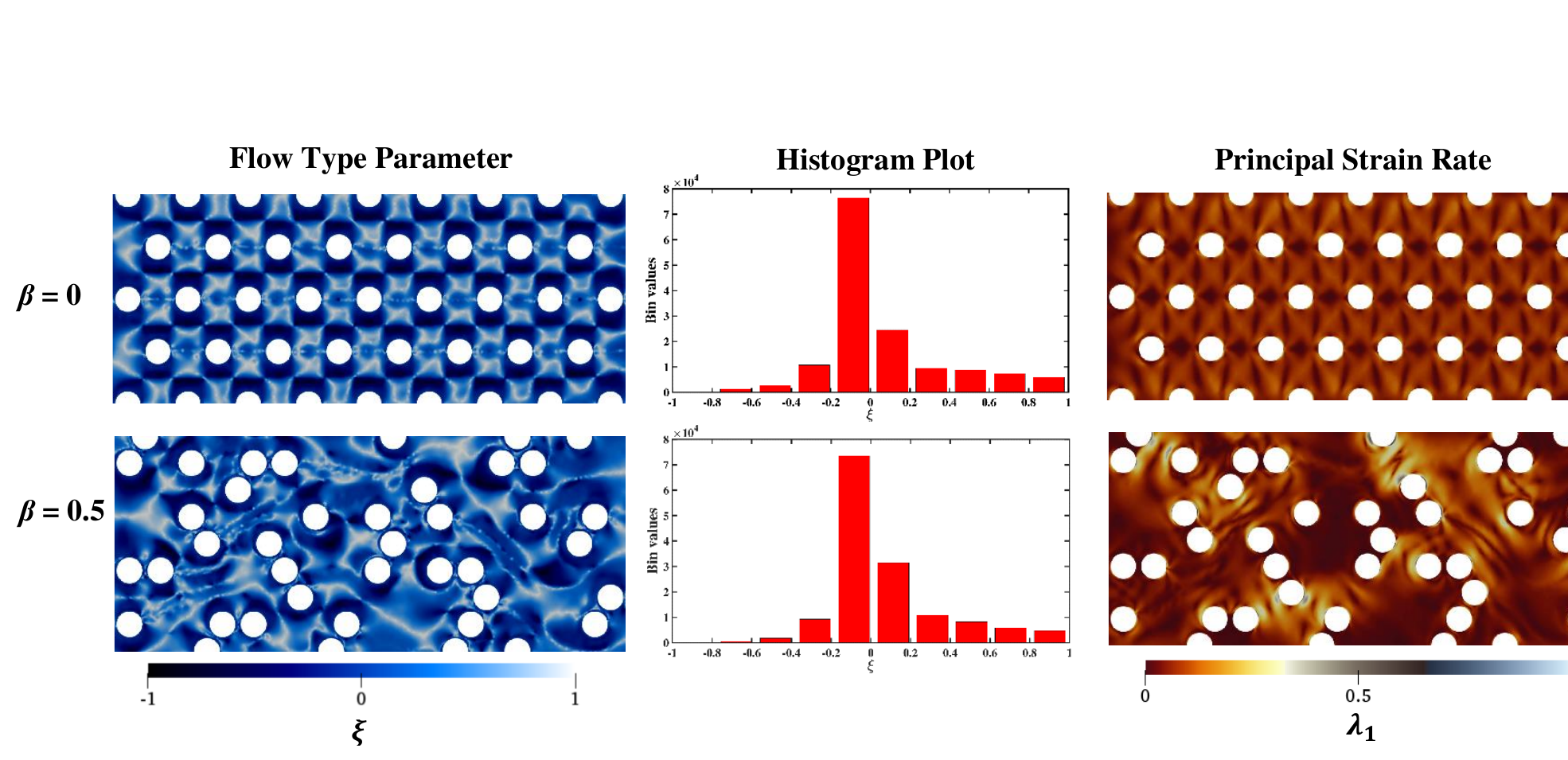}
    \caption{Surface and histogram plots of the flow type parameter. Variation of the principle strain rate with the disorder parameter. Here all the results are presented at $Wi  =10$.}
    \label{fig:fig6}
\end{figure}

To explain the chaotic flow dynamics, we have first plotted the extension of polymers (sum of the trace components of the conformation tensor of polymers) induced by the flow field as a function of the Weissenberg number and disorder parameter in figure~\ref{fig:fig5}. A region of high polymer extension is seen around the vicinity of the micropillars due to the presence of a strong shearing flow field. Furthermore, at the downstream of these micropillars, a region of high valued polymer extension is seen, also popularly known as the birefringent strand. This is because of the presence of a strong extensional flow field in this region, which results in a high stretching of the polymer molecules. One can see that the length, width, and the value of these birefringent strands increase as we increase both the Weissenberg number and disorder parameter. For an ordered and aligned configuration, these strands are not developed properly downstream of the micropillars in the flow direction (for instance, see the results presented for $Wi = 10$ and $\beta = 0$) as most of the fluids avoid these regions to flow, as shown in the velocity magnitude plot in figure~\ref{fig:fig2}. We can also say that the stagnation points of the micropillars are screened from the fluid flow, thereby resulting in less stretching of the polymer molecules. However, as the geometric disorder is gradually introduced in this ordered and aligned configuration, these stagnation points are now progressively revealed to the flow field, which cause an increase in the extensional flow strength downstream of the micropillars. This ultimately increases the polymer stretching. This is confirmed by calculating the flow type parameter $\xi$, which provides information about the nature of the local fluid deformation~\citep{walkama2020disorder,haward2021stagnation}. It is defined as $\xi = \frac{(||\textbf{D}|| - ||\boldsymbol{\Omega}||)}{(||\textbf{D}|| - ||\boldsymbol{\Omega}||)}$, where $\textbf{D}$ and $\boldsymbol{\Omega}$ are the deformation rate and vorticity tensors, respectively. A value of $\xi = -1$ signifies pure rotation, whereas the values of 0 and 1 quantify pure shear and extension, respectively. Figure~\ref{fig:fig6} shows both the surface and histogram plots of this flow type parameter at two values of the disorder parameter, namely, 0 and 0.5 at a fixed value of $Wi = 0$. Irrespective of the disorder parameter, it can be seen that the shear-dominated deformation region is mostly present in the flow field. However, as the disorder parameter increases, the shear-dominated region is decreased whereas the extension-dominated region is increased, as it is evident from the histogram plot. The flow strength of this extension-dominated region is evaluated by the principal strain rate $\lambda_{1} = \sqrt{(\text{D}_{11} - \text{D}_{22})^{2} + 4\text{D}_{12}^{2}}$, which is the eigenvector of $\textbf{D}$. One can clearly see that the strength of the extension-dominated region increases as we introduce some finite values of the geometric disorder in an initial ordered and aligned configuration of the micropillars. This causes greater stretching of the polymer molecules (as already shown in figure~\ref{fig:fig5}) in a disordered configuration of the micropillars as compared to that in an ordered one. As a result, the intensity of the chaotic fluctuations in the flow field is more in the former geometry than that in the latter one.

\section{\label{Con}Conclusions}
The flow of viscoelastic fluids through a porous media becomes chaotic and turbulent-like (the so-called elastic turbulence) once the flow rate exceeds a critical value. In this study, we have shown that these chaotic flows of viscoelastic fluids through a model porous media, consisting of a microchannel with many micropillars present in it, strongly depend on the geometric disorder in the arrangement of these micropillars. In particular, the present numerical study has shown that the introduction of geometric disorder into an initial ordered and aligned configuration of the micropillars increases the chaotic fluctuations in the flow field of viscoelastic fluids. This is in contrast to that seen in a recent experiment for an initial ordered and staggered configuration of the micropillars wherein the chaotic fluctuations are suppressed with the geometric disorder. Therefore, our results show that the geometric disorder does not always suppress the chaotic fluctuations in flows of viscoelastic fluids through a porous media. These chaotic fluctuations are actually dependent on the number of the stagnation points of these micropillars revealed to the flow field wherein maximum stretching of the viscoelastic microstructure happens. This gradual stretching of the microstructure leads to an increase of the elastic stresses, causing the occurrence of an initial elastic instability, which ultimately leads to the generation of chaotic and fluctuating flow fields with the increase in the flow rate. For an initial ordered aligned arrangement of the micropillars, the number of these stagnation points increases with the geometric disorder, whereas it decreases for an initial ordered staggered arrangement. Our findings based on the extensive numerical simulations are perfectly in line with a more recent experimental work.

\section{Acknowledgements}
Authors would like to thank IIT Ropar for providing funding through the ISIRD research grant (Establishment1/2018/IITRPR/921) to carry out this work. 

\section{Declaration of interests}
The authors report no conflict of interest

\bibliographystyle{jfm}
 Note the spaces between the initials
\bibliography{jfm-instructions}

\end{document}